\documentclass[aps,prc,reprint,superscriptaddress,nofootinbib]{revtex4-2}
\usepackage{amsmath}
\usepackage{amssymb}
\usepackage{graphicx}
\usepackage{dcolumn}
\usepackage{bm}
\usepackage{braket}
\usepackage{mathrsfs}
\usepackage{booktabs}
\usepackage{multirow}
\usepackage{url}
\usepackage{physics}
\usepackage{subfigure}
\usepackage[colorlinks=true,linkcolor=blue, filecolor=blue,urlcolor=blue,citecolor=blue]{hyperref}
\usepackage{color}
\usepackage{algorithm}      
\usepackage{algorithmicx}   
\usepackage{algpseudocode}  

\begin{document}
	\title
	{Studying few cluster resonances  with quantum neural network driven iterative Harrow-Hassidim-Lloyd  algorithm }
	
	\author{Hantao Zhang}
	\email{zhang\_hantao@foxmail.com}
	\affiliation{School of Physics Science and Engineering, Tongji University, Shanghai 200092, China}
	
	\author{Dong Bai}
	\email{dbai@hhu.edu.cn}
	\affiliation{College of Mechanics and Engineering Science, Hohai University, Nanjing 211100, Jiangsu, China}
	
	\author{Zhongzhou Ren}
	\email[Corresponding author: ]{zren@tongji.edu.cn}
	\affiliation{School of Physics Science and Engineering, Tongji University, Shanghai 200092, China}
	\affiliation{Key Laboratory of Advanced Micro-Structure Materials, Ministry of Education, Shanghai 200092, China}

\begin{abstract}
By using the quantum computing  the  properties of hypernuclei ${}^5_{\Lambda}$He, ${}^{\  6}_{{\Lambda\Lambda}}$He and ${}^9_{\Lambda}$Be can be investigated within microscopic cluster model. Our approach combines quantum neural network (QNN) with  iterative Harrow-Hassidim-Lloyd (IHHL)  algorithm (abbreviated as QNN-IHHL) to solve the quantum many-body problem. To efficiently describe resonance phenomena, we employ complex scaling and eigenvector continuation techniques, providing a robust framework for identifying few-cluster resonance parameters within quantum computing. To validate our quantum algorithm, the resonant $4^{+}$  state of ${}^9_{\Lambda}$Be is chosen as a core example. With QNN-IHHL algorithm we realize a fully quantum workflow, which provides a novel framework and some ground work for exploring resonance properties in complex nuclear many-body systems.

\end{abstract}

\maketitle

%
%
%













%
%
%
%

%
%
%
%
%
%
%
%
\section{introduction}
{\color{black}Quantum computing, situated at the intersection of quantum mechanics and computer science, is rapidly transforming our ability to address problems long considered intractable on classical machines. By exploiting fundamental principles such as quantum superposition and entanglement, quantum processors offer access to an exponentially larger state space, enabling new strategies for simulating quantum many-body systems. Applications of this paradigm span across chemistry, condensed matter, high-energy physics, and nuclear physics, where the quantum nature of the systems under study aligns naturally with the computational framework.

In nuclear physics, the exploration of entanglement has emerged as a powerful diagnostic tool for understanding the structure and dynamics of hadrons and nuclei \cite{Savage:2023qop,Klco_2022,Ho_2016,PhysRevD.95.114008,PhysRevD.98.054007,PhysRevLett.122.102001,Tu_2020,BEANE2021168581,ISKANDER2020135948,Kruppa_2021,doi:10.1142/S0217751X21502055,PhysRevD.104.L031503,PhysRevC.106.024303,PhysRevC.103.034325,PhysRevD.104.074014,PhysRevD.106.L031501,Bai:2022hfv,Johnson_2023,EHLERS2023169290,Tichai_2023,Pazy_2023,Bulgac_2023,PhysRevA.105.062449,PhysRevC.107.044318,PhysRevC.105.014307,Bai:2023rkc,Robin_2023,Gu_2023,Sun_2023,Bai:2023tey,Bai:2023hrz,Bai:2024omg,miller2023entanglementmaximizationlowenergyneutronproton,Hengstenberg_2023,P_rez_Obiol_2023,Gorton_2024}. Recent studies have demonstrated how entanglement measures can reveal spatial correlations, identify emergent degrees of freedom, and provide insight into the nature of confinement and clustering in strongly interacting systems. These developments have motivated the application of quantum information methods, not merely as computational tools, but as conceptual instruments to probe fundamental properties of matter. Quantum computing offers transformative potential for nuclear physics by efficiently solving quantum many-body problems and simulating high-dimensional state evolution - capabilities particularly valuable for nuclear systems. Recent applications to nuclear reactions and structure problems  demonstrate its growing role in tackling previously intractable challenges 	\cite{Roggero_2019,Mueller_2020,Turro_2023,Baroni_2022,bedaque2022radiativeprocessesquantumcomputer,turro2024evaluationphaseshiftsnonrelativistic,Du_2021,Du:2024zvr,Parnes:2025seu}.
	
Quantum algorithms designed for eigenvalue problems, such as the variational quantum eigensolver (VQE)\cite{osti_1623945,Peruzzo_2014,McClean_2016,PhysRevLett.120.210501,RevModPhys.94.015004,PhysRevC.104.024305} and its various variants \cite{Higgott_2019,McArdle_2019,Yuan_2019,Grimsley_2019,PhysRevResearch.2.043140,Stokes_2020,Gomes_2021,PRXQuantum.2.020310,PhysRevC.105.064317,Koczor_2022} and quantum phase estimation (QPE) \cite{kitaev1995quantummeasurementsabelianstabilizer,Nielsen_Chuang_2010}, have already demonstrated success in solving electronic structures and simulating small quantum systems. Their adaptation to nuclear physics has enabled progress in modeling nuclear potentials, evaluating scattering amplitudes, and extracting low-lying spectra from ab initio interactions. Extensions of these algorithms, including ansatz-efficient VQE variants and hybrid quantum-classical schemes, have pushed further toward describing realistic nuclei and nuclear reactions. More recently, quantum neural networks (QNN) 	\cite{Cong_2019,Beer_2020,Abbas_2021,Pan_2023,Jin_2024}  have emerged as a versatile class of models capable of learning complex quantum correlations, representing non-trivial quantum states, and enabling efficient encoding of physical priors.		
	
Despite these advances, several frontiers remain unresolved. One of the most important challenges lies in the quantum treatment of non-Hermitian systems. Resonant states—complex-energy solutions corresponding to metastable configurations—play a central role in nuclear scattering, decay processes, compound nucleus formation, and the determination of reaction cross sections. The non-Hermitian character of resonances complicates their representation on quantum hardware, where unitary evolution is typically assumed. Developing scalable, noise-resilient quantum algorithms that can faithfully capture resonance phenomena is therefore an open and urgent problem, with wide-ranging implications for nuclear structure, astrophysics, and so on.	
In our previous work, we have conducted some studies on non-Hermitian nuclear resonant states for two body system
\cite{Zhang:2024rpa} and proposed a novel iterative non-Hermitian algorithm 
based on the Harrow-Hassidim-Lloyd algorithm \cite{PhysRevLett.103.150502} for solving bound state and exacting resonances with complex scaling method  \cite{Aguilar1971ACO,Balslev1971SpectralPO,10.1143/PTP.116.1,Myo:2014ypa,PhysRevC.89.034322,Zhang:2022rfa,Zhang:2023dzn,Myo:2023btg,Zhang:2024ril,Zhang:2024gac,10.1088/1674-1137/ad88fa,10.1088/1674-1137/ad9a8c} and eigenvector continuation (EC) \cite{PhysRevLett.121.032501,PhysRevC.107.064316,PhysRevC.101.041302,KONIG2020135814,COMPANYSFRANZKE2022137101,10.1093/ptep/ptac057,PhysRevLett.126.032501}, which has already  found many applications in low energy nuclear physics and in particular, it has
been utilized to build emulators in the context of nuclear scattering \cite{Melendez_2022,PhysRevC.106.054322,10.3389/fphy.2022.1092931,FURNSTAHL2020135719,DRISCHLER2021136777,PhysRevC.103.014612,PhysRevC.106.024611,PhysRevC.107.054001}.

{\color{black} In this work, we present an enhanced implementation of the iterative Harrow-Hassidim-Lloyd (IHHL) algorithm \cite{zhang2025iterativeharrowhassidimlloydquantumalgorithm} through combination with quantum neural network (QNN), establishing a full process quantum computational framework QNN-IHHL for solving three-cluster resonance problems.}
{\color{black} QNN offers a promising framework for modeling complex interactions in nuclear physics by leveraging the expressive power of quantum states. Their ability to naturally represent high-dimensional Hilbert spaces makes them well-suited for simulating many-body wave functions, scattering amplitudes, and reconstructing nuclear states. 
}
%
To validate the reliability of our QNN-IHHL approach, we perform quantum simulations to compute the resonant state of the ${}^9_{\Lambda}$Be system. Our new algorithm also provide possibilities in applying the bound state technique such as trap method \cite{Zhang:2019cai,Zhang:2020rhz,Guo:2021uig,Guo:2021qfu,Zhang:2024vmz,Zhang:2024mot,Zhang:2024vch,Zhang:2024ykg,Guo:2025ngh} to handle scattering phase shifts. 

The rest parts are organized as follows: In Sec.II, we introduce the framework of quantum neural network, iterative Harrow-Hassidim-Lloyd algorithm and eigenvector continuation with complex scaling. In Sec.III, the numerical results  of resonance in ${}^9_{\Lambda}$Be system are presented and discussed. Sec.IV summarizes the article. Some key data and numerical details are listed in the appendix.

}

\section{theoretical formalism}
\label{Theoretical Formalism}

\subsection{Microscopic three cluster model for ${}^{5}_{{\Lambda}}$He, ${}^{\  6}_{{\Lambda\Lambda}}$He and ${}^{9}_{{\Lambda}}$Be}
The total wave function of ${}^{5}_{{\Lambda}}$He, ${}^{\  6}_{{\Lambda\Lambda}}$He and ${}^{9}_{{\Lambda}}$Be can be formulated as:
\begin{equation}
	\label{config.equ}
	\begin{aligned}
		&\Psi_J({}^5_{\Lambda}\mathrm{He})=\sum_{d}f_J(d)\phi_{\alpha}\phi_J(d,r)Y_J(\hat{\textbf{r}}), 
		\\
&\Psi_J^{}({}^{\ 6}_{\Lambda\Lambda}\mathrm{He})=\sum_{dD,lL}f_{lLJ}(d,D)[\phi_{\alpha}\phi^{\Lambda\Lambda}_{l}(d,{r})Y_l(\hat{\bm{r}})\\&\times\phi^{\alpha-\Lambda\Lambda}_L(D,R)Y_L(\hat{\bm{R}})]_J,\ \\
&\Psi_J^{}({}^9_{\Lambda}\mathrm{Be})=\sum_{lL}\sum_{dD}f_{lLJ}(d,D)[\Phi_{l,\alpha\alpha}(d)\phi_L(D,R)Y_L(\hat{\textbf{R}})]_J,
	\end{aligned}
\end{equation}
where $\phi_{\alpha}$ represents the wave functions of $\alpha$ particle, the wave function $\phi^{c_1-c_2}_{l}({R_{1,2}},{r_{1,2}})$ is taken as the form of local Gaussian function,
\begin{equation}
	\label{config_equ}
	\begin{aligned}
	&\phi^{c_1-c_2}_l({R_{1,2}},{r_{1,2}})= (4\pi)(\sqrt{\pi} b_{c_1,c_2})^{-3/2} \exp\left[ -\frac{R^2 + D^2}{2b_{c_1,c_2}^2} \right] \\&\times \mathscr{J}_i \left( \frac{DR}{b_{c_1,c_2}^2} \right),
	\end{aligned}
\end{equation}
where the size parameters $b$ are defined by Eq. (\ref{b_parameter}) and function $\mathscr{J}(x)$ is defined by modified Bessel function of the first kind $I$,
\begin{equation}
	\begin{aligned}
		& \mathscr{J}(x)=\sqrt{\dfrac{\pi}{2x}}I_{L+1/2}(x).
	\end{aligned}
\end{equation}

\begin{figure}[htbp] 
	\centering
	{\includegraphics[width=0.5\textwidth]{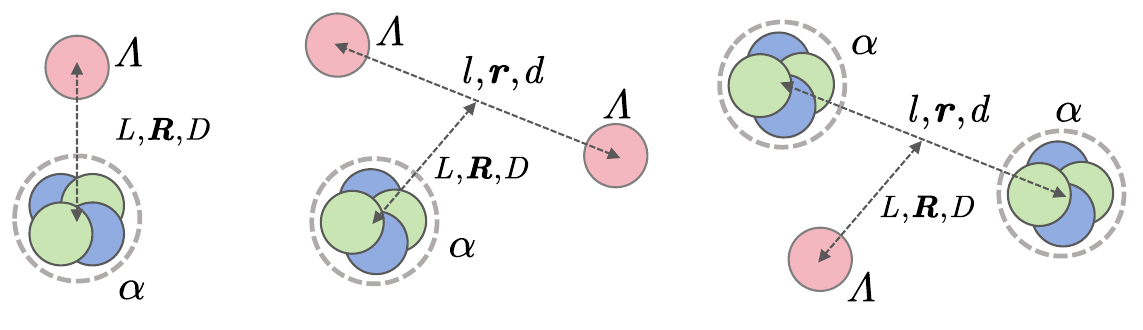} }
	\caption{Coordinate system and angular momentum labels for ${}^{5}_{{\Lambda}}$He, ${}^{\ 6}_{{\Lambda \Lambda}}$He and ${}^{9}_{{\Lambda}}$Be.}	\label{three_system} 
\end{figure}

 For the coordinates in Fig.\ref{three_system} the appropriate size parameters $b$ are chosen so that combinations with the  corresponding reduced massed lead to an identical value of $\hbar/\omega$:
\begin{equation}
	\label{b_parameter}
	\begin{aligned}
		&M_Nb_N^2=M_{\Lambda}b_{\Lambda}^2=\mu_{\alpha \alpha}b_{\alpha \alpha}^2=\mu_{\Lambda \alpha}b_{\Lambda \alpha}^2=\mu_{\Lambda, \alpha\alpha}b_{\Lambda,\alpha\alpha}^2\\&=\mu_{\Lambda\alpha, \alpha}b_{\Lambda\alpha, \alpha}^2=\hbar/\omega,
	\end{aligned}
\end{equation}
where $b_N$ is taken as 1.36 fm for $\alpha$ particle.
{\color{black} By adjusting parameter $b_N$, we can achieve a better description of the ground-state energies for these three hypernuclei. However, to facilitate comparison with our earlier calculations \cite{Zhang:2025mrl}, we retain the standard value $b_N$ = 1.36 fm. 
	

The Hamiltonian of ${}^{5}_{{\Lambda}}$He can be written as
\begin{equation}
	\begin{aligned}
		&
		H({}^{5}_{{\Lambda}}He)=H_{\alpha}+H_R^{},
	\end{aligned}
\end{equation}
with
\begin{equation}
	\begin{aligned}
		&H_R^{}=T_R^{}+\sum_{i=1}^4V_{\Lambda N}^i.
	\end{aligned}
\end{equation}

The total Hamiltonian of ${}^{\  6}_{{\Lambda\Lambda}}$He and ${}^{9}_{{\Lambda}}$Be can be divided as:
\begin{equation}
	\begin{aligned}
		&
		H({}^{\  6}_{{\Lambda\Lambda}}He)=H_{\alpha}+H_R^{}+H_{r},
	\end{aligned}
\end{equation}
with
\begin{equation}
	\begin{aligned}
&H_R^{}=T_R^{}+\sum_{i=1}^4V_{\Lambda N}^i,\  H_{r}=T_{r}+V_{\Lambda\Lambda},
	\end{aligned}
\end{equation}
and
\begin{equation}
	\begin{aligned}
		&
		H({}^{9}_{{\Lambda}}Be)=H_{{}^8_{}Be}+H_R^{},
	\end{aligned}
\end{equation}
with
\begin{equation}
	\begin{aligned}
		&  H_{{}^8_{}Be}=T-T_G+V_N+V_C,\ \\&H_R^{}=T_R^{}+\sum_{i=1}^8V_{\Lambda N}^i,
	\end{aligned}
\end{equation}
where $T_R$ is the  kinetic energy associated with the coordinate $R$ and $T_r$ is the  kinetic energy associated with the coordinate $r$. $V_{\Lambda N}^i$ denotes the $\Lambda N$ interaction between the hyperon $\Lambda$ and the $i$-th nucleon. $V_{\Lambda\Lambda}$ is the  interaction  between two $\Lambda$ hyperon.

{\color{black}
For effective two-body nuclear potential we adopt the Volkov interaction:
\begin{equation}
	\begin{aligned}
		&V_N=\dfrac{1}{2}\sum\limits_{i\neq{j}}^8\{(W-MP_{\sigma\tau}+BP_{\sigma}-HP_{\tau})\}_{ij}\\&\times\sum\limits_{k=1}^{k_{max}}V^0_{k} 
		\exp(-\dfrac{r_{ij}^2}{a_k^2}),
	\end{aligned}
\end{equation}
where the parameters of Volkov and Minnesota interactions can be found in \cite{Zhang:2025mrl}.  {\color{black} In our calculations Volkov No.1 is utilized as the $NN$ interaction.}

The Coulomb interaction can be written as
\begin{equation}
	\begin{aligned}
		V_C=\dfrac{1}{2}\sum\limits_{i\neq{j}}^8(\dfrac{1}{2}+t_{zi})(\dfrac{1}{2}+t_{zj})\dfrac{e^2}{r_{ij}},
	\end{aligned}
\end{equation}
where the isospin z-component equals $t_z=+{1}/{2}$ for the proton and $t_z=-{1}/{2}$ for the neutron.
}

The two-body $\Lambda N$ interaction is chosen 
as the YNG interaction
, which is given by:
\begin{equation}
	\begin{aligned}
		&V_{\Lambda N}(r)=\sum_i\{(V_D^0+V_{EX}^0 P_r)\exp[-(\dfrac{r}{\beta_{i}})^2]\\&+(V_{D}^{\sigma\sigma}\sigma_{\Lambda}\sigma_{N} +V_{EX}^{\sigma\sigma}\sigma_{\Lambda}\sigma_{N} P_r)\exp[-(\dfrac{r}{\beta_{i}})^2]\},
	\end{aligned}
\end{equation}
where $P_r$ is the space exchange parameter and 
\begin{equation}
	\begin{aligned}
		\begin{cases}
			&V_D^{0}=\dfrac{V^0(E)+V^0(O)}{2},\ 
			V_{EX}^{0}=\dfrac{V^0(E)-V^0(O)}{2}\\
			\\~
			&V_{D}^{\sigma\sigma}=\dfrac{V^0_{\sigma\sigma}(E)+V^0_{\sigma\sigma}(O)}{2},\ 
			V_{EX}^{\sigma\sigma}=\dfrac{V^0_{\sigma\sigma}(E)-V^0_{\sigma\sigma}(O)}{2}.
		\end{cases}
	\end{aligned}
\end{equation}
The parameters of the YNG model  used in this work are taken from \cite{10.1143/PTPS.81.42} and listed in Table. With the parameters adopted in $\Lambda N$ potential, the binding energy of  ${}^{5}_{\Lambda}$He can be reproduced as 3.10 MeV, which is in good agreement with the experimental value
3.12$\pm$0.02 MeV \cite{RevModPhys.88.035004}. 
\begin{table}[h]
	\centering
	\caption{$\Lambda N$ interaction depth of the YNG model. The Fermi momentum $k_F$ is 0.9 fm$^{-1}$.The unit of $V^0$ is MeV and the unit of $\beta_i$ is fm.}
	\label{tab:YNG_interaction}
	\begin{tabular}{ccccc}
		\hline
			\hline
		$\beta_i$ & $V^0(E)$ & $V^0(O)$ & $V^0_D$ & $V^0_{EX}$ \\
		\hline
		1.5 & -9.93  & -7.66   & -8.795   & -1.135  \\
		0.9 & -227.73 & -82.55  & -155.140 & -72.590 \\
		0.5 & 1021.17 & 717.40  & 869.285  & 151.885 \\
		\hline
			\hline
	\end{tabular}
\end{table}

\begin{table}[h]
	\centering
	\caption{OBE-simulating $\Lambda\Lambda$ potential. Depths $v_0^{i}$ and $v_{\sigma\sigma}^{i}$ (MeV) for each range $\beta^{i}_{\Lambda\Lambda}$ (fm).}
	\label{tab:OBE_LL_potential}
	\begin{tabular}{cccc}
		\hline
		\hline
		$\beta^{i}_{\Lambda\Lambda}$  & 1.342 & 0.777 & 0.35 \\
		\hline
		$v_0^{i}$  & -21.34 & -187.0 & 10850 \\
		$v_{\sigma\sigma}^{i}$  & 0.1932 & 32.17 & 2035 \\
		\hline
		\hline
	\end{tabular}
\end{table}

The two-body $\Lambda-\Lambda$ interaction is chosen as the following three range  Gaussian form:
\begin{equation}
	\begin{aligned}
		&V_{\Lambda\Lambda}(r)=\sum_{i}(v^i_{0}+v^{i}_{\sigma\sigma}\bm{\sigma}_{\Lambda}\cdot\bm{\sigma}_{\Lambda})\exp[-(\dfrac{r}{\beta^{i}_{\Lambda\Lambda}})^2],
	\end{aligned}
\end{equation}
where parameters $\beta_{\Lambda\Lambda}$, $v_0$ and $v_{\sigma\sigma}$ are taken from \cite{Hiyama:1997ub} and listed in Table . With the parameters adopted in $\Lambda N$ and $\Lambda \Lambda$ interactions, the binding energy of  ${}^{\  6}_{{\Lambda\Lambda}}$He can be reproduced as 6.6 MeV, which is in good agreement with the experimental value
6.91$\pm$0.16 MeV \cite{PhysRevC.88.014003} and the binding energy of  ${}^{9}_{{\Lambda}}$Be can be reproduced as 6.99 MeV, which is also in good agreement with the experimental value
6.63 MeV \cite{BERTINI1981365,PhysRevLett.51.2085,BRUCKNER1976481}. 

\begin{figure*}[htbp] 
	\centering
	{\includegraphics[width=1\textwidth]{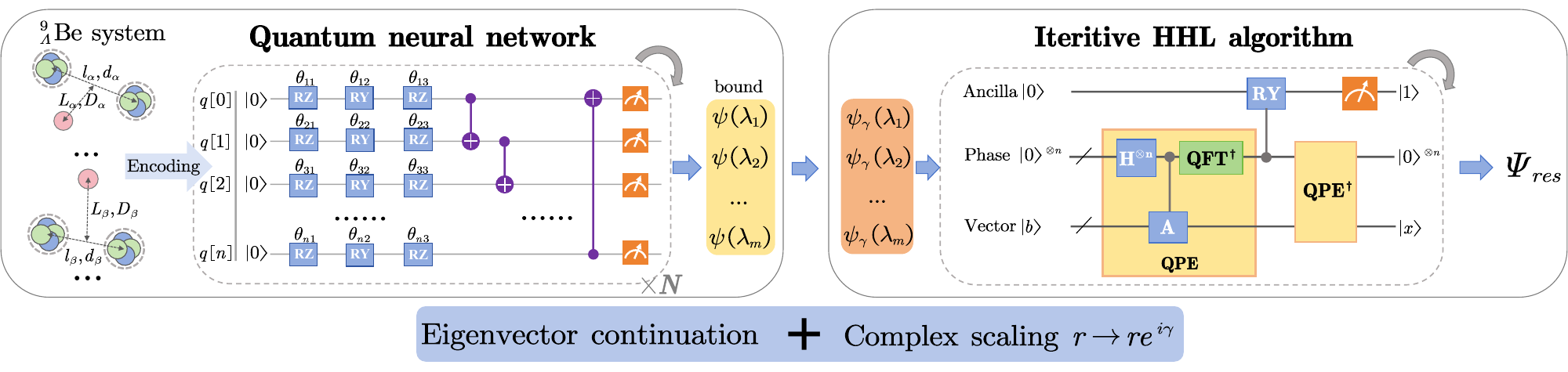} }
	\caption{The gray arrows in the upper right corner of QNN and IHHL algorithm indicate the need for iteration.}	\label{QC_alpha_2Lambda_algorithm} 
\end{figure*}

\subsection{Quantum neural network}
When selecting a suitable set of  orthonormal basis functions $\{\phi_i\}$, the parameterized Hamiltonian $H(\lambda)$ can be expressed in the following form with creation and annihilation operators,
\begin{equation}
	\begin{aligned}\label{}
		&H(\lambda)=\sum_{i,j}^{}h_{i,j}(\lambda)a_i^{\dagger}a_j^{},\ \ \ h_{i,j}(\lambda)=\bra{\phi_i}{T+V_N(\lambda)+V_C}\ket{\phi_j},
	\end{aligned}
\end{equation}
similarly, if the basis functions are not orthonormal the overlap operator $N$ can be written as,
\begin{equation}
	\begin{aligned}\label{}
		&N=\sum_{i,j}^{}n_{i,j}(\lambda)a_i^{\dagger}a_j^{},\ \ \ n_{i,j}(\lambda)=\bra{\phi_i}{N}\ket{\phi_j}.
	\end{aligned}
\end{equation}

As the first step to utilize quantum neural network (QNN) to solve the three-cluster bound state problem, the three cluster Hamiltonian and overlap operator should be  encoded  into a qubit representation
through the Jordan-Wigner transformation \cite{Jordan1928berDP}, with which the fermionic creation and annihilation operators can be expressed as,
\begin{equation}
	\begin{aligned}\label{}
		&a_j^{\dagger}=\dfrac{1}{2}(X_j-iY_j)\otimes Z_{j-1}^{\mathscr{D}},\\
		&a_j^{}=\dfrac{1}{2}(X_j+iY_j)\otimes Z_{j-1}^{\mathscr{D}},
	\end{aligned}
\end{equation}
where $X$, $Y$, and $Z$ are Pauli operators, $Z_{j-1}^{\mathscr{D}}$  is defined as
\begin{equation}
	\begin{aligned}\label{}
		&Z_{j-1}^{\mathscr{D}}=Z_{j-1}\otimes Z_{j-2}\otimes \cdots \otimes Z_0.
	\end{aligned}
\end{equation}

In addition to the Jordan-Wigner transformation, there are other quantum mapping methods. For example, for $2^{m}\times2^{m}$ matrices, one can choose to represent the Hamiltonian matrix using $m$ qubits, which can effectively reduce the number of required qubits,

\begin{equation}
	\begin{aligned}\label{}
		&H(\lambda)=\sum_{i_1,i_2,\cdots,i_m=0,1,2,3}c_{i_1,i_2,\cdots,i_m}(\lambda)(\sigma_{i_1}\otimes\sigma_{i_2}\otimes\cdots\otimes\sigma_{i_m}),
	\end{aligned}
\end{equation}
where
\begin{equation}
	\begin{aligned}\label{}
		&c_{i_1,i_2,\cdots,i_m}(\lambda)=\dfrac{1}{2^{m}}Tr((\sigma_{i_1}\otimes\sigma_{i_2}\otimes\cdots\otimes\sigma_{i_m})H(\lambda)),\\& \ \sigma=\{I,X,Y,Z\}.
	\end{aligned}
\end{equation}

The core of quantum neural network is the parameterized strongly entangled circuit, which can be functionally analogous to the hidden layers composed of neurons in traditional neural networks. Based on the variational principle, we can obtain the ground state of a bound system by minimizing the energy $E$ as loss function $\mathcal{L}$. The energy expectation value of the operator 
$H$ can be expressed as a function of the parameters $\theta$ of the QNN circuit,
\begin{equation}
	\begin{aligned}\label{}
		&E(\theta)=\bra{0}U^{\dagger}(\theta)HU(\theta)\ket{0},
	\end{aligned}
\end{equation}
where $U$ is the quantum state generated by QNN, note that if the basis vectors are non-orthogonal, the energy expectation value becomes,
\begin{equation}
	\begin{aligned}\label{}
		&E(\theta)=\dfrac{\bra{0}U^{\dagger}(\theta)HU(\theta)\ket{0}}{\bra{0}U^{\dagger}(\theta)NU(\theta)\ket{0}}.
	\end{aligned}
\end{equation}

Similar to improving performance of the network by adding more hidden layers in traditional neural networks, in QNN the parameterized circuit shown in Fig. \ref{QC_alpha_2Lambda_algorithm} will be repeated to increase circuit depth. In some cases (and also in this work), slight modifications are made to the repeated layers, such as cyclically altering the connection sequence of the CNOT gates. If we denote the quantum gate with control bit $q[i]$ and target bit $q[j]$  as CNOT$(i,j)$, then the connections we use in this work can be expressed as: CNOT$(1,i_1)$, CNOT$(2,i_2)$, ..., CNOT$(N,i_N)$, where $\{i\} = \{2, 3, \cdots, N-1, N, 1\}, \{3, 4, \cdots, N, 1, 2\}, \cdots $ for each parameterized circuit module. 


Using the so-called parameter shift rule \cite{Wierichs_2022}, the partial derivative of energy with respect to the parameter $\theta$ can be obtained,
\begin{equation}
	\begin{aligned}\label{}
		&\nabla_{\theta} E(\theta)=\dfrac{1}{2}[E(\theta+\dfrac{\pi}{2})-E(\theta-\dfrac{\pi}{2})],
	\end{aligned}
\end{equation}
which will be used in the optimizer for updating the parameters in the quantum gates. For the choice of optimizer,  we can employ various approaches, including stochastic gradient descent (SGD) \cite{SGDWIKI} and  alternative approaches such as the Adam  optimizer \cite{SGDWIKI} and its variants, or gradient-free methods like Rotosolve method \cite{pankkonen2025improvingvariationalquantumcircuit} may also be implemented. For our current calculations in this work the basic version of SGD is adopted:
	\begin{equation}
	\theta_{t+1} = \theta_t - \eta \cdot \nabla_\theta \mathcal{L}(\theta_t),
\end{equation}
of course we can also consider other variant  of SGD  with such as  momentum, Nesterov approach \cite{SGDWIKI} and so on. {\color{black}Here, $t$ and $t+1$ denote the current and subsequent time steps, respectively, and $\eta$ is the learning rate which can be modified during the QNN iteration process.}


To solve for excited state energy levels above the ground state using QNN, one can employ a projection method to incorporate a pseudo-potential into the original Hamiltonian,
\begin{equation}
	\begin{aligned}    		
		H_{p}=H+c\sum_{i}\ket{\phi^{l}_i}\bra{\phi^{l}_i},
	\end{aligned}
\end{equation}
where $\phi^{l}_i$ are the lower eigenstates and  $c$ is a sufficiently large constant to ensure that the variational energy can reach the excited state.

\begin{figure*}[htbp] 
	\centering
	{\includegraphics[width=1\textwidth]{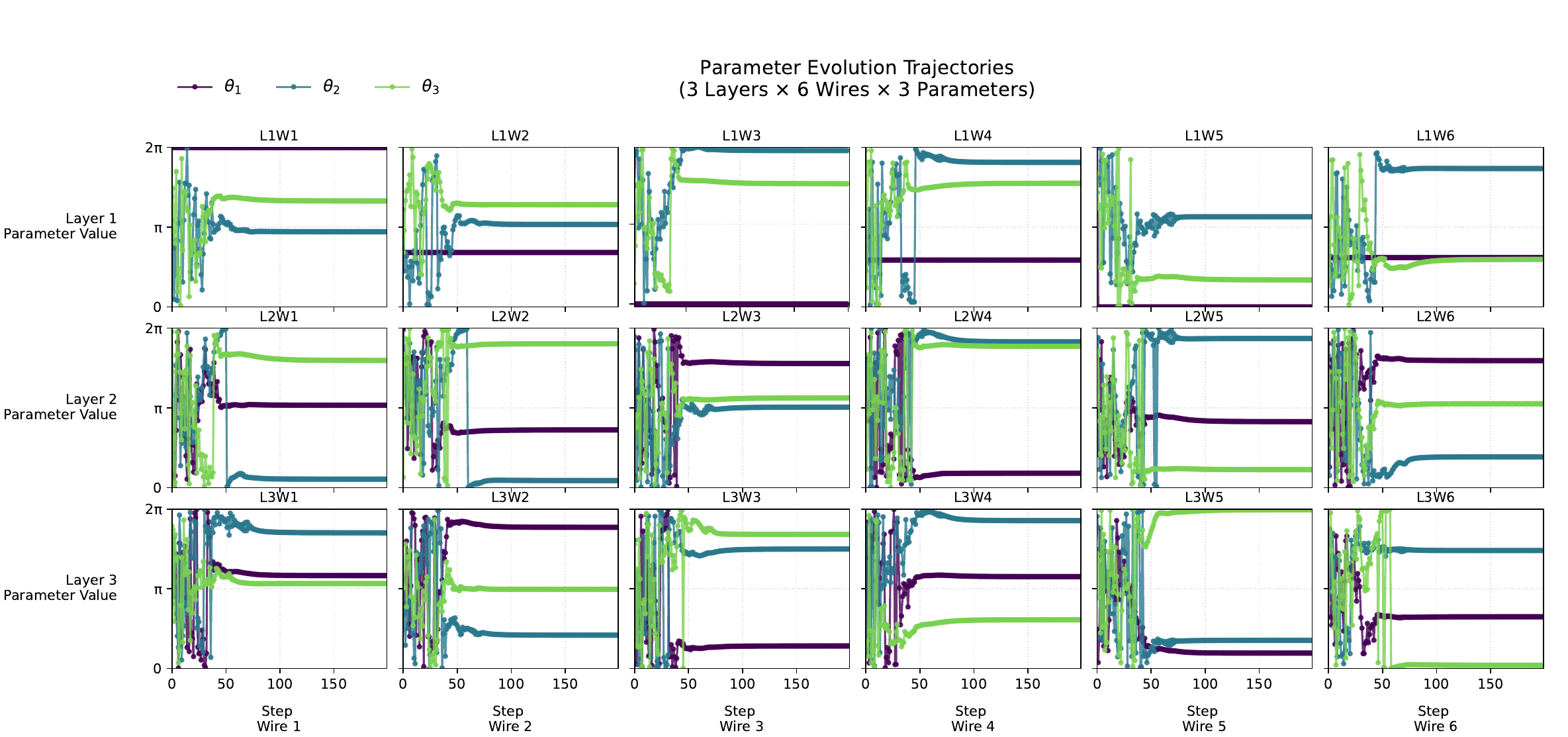} }
	\caption{Each subplot displays the angles $\theta_1$, $\theta_2$, and $\theta_3$, corresponding to the rotation angles of the $RZ$, $RY$, and $RZ$ gates (normalized to $[0, 2\pi]$) in a unit, respectively. The horizontal axis represents three distinct quantum circuit layers, while the vertical axis denotes six different qubits. After 200 iterations of this $3\times6\times3$ QNN circuit, the rotation gate parameters converge to stable values.}	\label{gate_parameter_QNN} 
\end{figure*}

\subsection{Iterative Harrow-Hassidim-Lloyd (IHHL) algorithm}

The Harrow-Hassidim-Lloyd algorithm is a pure quantum algorithm designed to solve  linear equation $Ax=b$, where $A$ is a Hermitian operator. As displayed in Fig.\ref{QC_alpha_2Lambda_algorithm}, it leverages quantum phase estimation (QPE) to find the eigenvalues of the matrix $A$, then uses quantum gates to compute the inverse of these eigenvalues. Finally, the algorithm reconstructs the solution vector $x$ in quantum form. HHL algorithm can provide exponential speedup over classical methods for Hermitian matrices. 
Building upon the HHL algorithm, we have proposed an iterative method for solving eigenvalue problems (including bound states and resonant states) \cite{zhang2025iterativeharrowhassidimlloydquantumalgorithm}.

According to the ABC theorem \cite{Aguilar1971ACO,Balslev1971SpectralPO}, the  two-body Hamiltonian $H({\lambda})$ will be transformed into a complex scaled one, $H^{\gamma}(\lambda)=e^{-2i\gamma}T+V_N(\lambda,re^{i\gamma})+e^{-i\gamma}V_C$. {\color{black}While for microscopic cluster model the complex scaling  can be implemented by complex scaled generator coordinates \cite{Zhang:2022rfa,Zhang:2023dzn,Zhang:2025mrl}.}

In order to address the eigenvalue problem for resonant states, we introduce an iterative method, specifically we can reformulate the Schrödinger equation, so that the final eigenvector becomes the fixed point of the equation as follows,
\begin{equation}
	\begin{aligned}\label{}
		&C(E,\beta)\phi=\phi,
	\end{aligned}
\end{equation}
where  matrix elements of $C$ operator is defined by
\begin{equation}
	\begin{aligned}\label{}
		&C_{ij}(E,\beta)=(\phi|\dfrac{e^{-2i\gamma}T+V_N(\lambda,re^{i\gamma})+e^{-i\gamma}V_C-(E-\beta)}{\beta}|\phi),
	\end{aligned}
\end{equation}
where parentheses $"()"$ denotes the $c$-product and $\beta\neq0$.

In order to utilize the HHL algorithm to solve resonances,  the non-Hermitian operator $C(E,\beta)$ should be extended to be a larger Hermitian matrix $A$ (assuming the basis functions are orthonormal), 
\begin{equation}
	\begin{aligned}\label{LargerM}
		&A=
		\begin{pmatrix}
			0& C(E,\beta) \\
			C^{\dagger}(E,\beta)&0	
		\end{pmatrix}
	.
	\end{aligned}
\end{equation}

After the above processing, we can utilize the HHL algorithm to iteratively solve for the eigenenergies of single channel resonant state. It is also obvious that the IHHL algorithm is suitable for coupled-channel calculations to handle nuclear structure and reaction problems.  The HHL algorithm solves for the unknown vector $x$, namely corresponding to the new complex wave function. The $c$-product of the new and old wave functions is then used to calculate the new complex energy, afterwards by setting a error tolerance $\varepsilon$, we can determine whether to continue iterating with the HHL algorithm until the desired accuracy is reached, and eventually output the convergent eigenvalue and eigenvector of the complex scaled Hamiltonian. A more detailed illustration about IHHL algorithm can be found in our previous work \cite{zhang2025iterativeharrowhassidimlloydquantumalgorithm}.

{\color{black}
Last but not least, the IHHL algorithm can in fact handle generalized eigenvalue problems by computing the inverse of the overlap matrix and  thereby reducing the problem to a standard eigenvalue problem. In the following cluster calculations for hypernuclei, we adopt the aforementioned approach to solve the Hill-Wheeler equation.
%
%
%
}

\subsection{Eigenvector continuation}

Eigenvector continuation is an approach that can extract the eigenstates $\phi(\lambda_{\odot})$ of a Hamiltonian $H(\lambda_{\odot})$ with target  parameter $\lambda_{\odot}$ by obtaining eigenstates $\phi_T=\{\phi(\lambda_{i})\}$ of the Hamiltonian $\{H(\lambda_i)\}$. Explicitly speaking,  the parameter set $\{\lambda_i\}$ is served as training data and the corresponding training eigenvectors will be utilized to form a new basis to solve the eigenvalues and eigenvectors of target Hamiltonian $H(\lambda_{\odot})$.
Therefore eigenvector continuation can generally reduces the dimension of the eigenvalue problem from a large Hilbert
space to a smaller subspace spanned by the training eigenvectors $\phi_T$. In this way the computational cost
for each target parameter can be significantly reduced.

One of the  fundamental applications of eigenvector continuation is to estimate bound states from bound states.  Specifically, for extracting bound states the generalized eigenvalue
problem should be solved,
\begin{equation}
	\begin{aligned}\label{}
		&H^{EC}\ket{\phi(\lambda_\odot)}=E(\lambda_{\odot})N^{EC}\ket{\phi(\lambda_\odot)},
	\end{aligned}
\end{equation}
where the Hamiltonian and overlap matrix elements $H(\lambda_{\odot})^{EC}$, $N(\lambda_{\odot})^{EC}$ are constructed by training eigenvectors of bound states, respectively,
\begin{equation}
	\begin{aligned}\label{}
		&H^{EC}_{ij}=\bra{\phi(\lambda_{i})}{H(\lambda_{\odot})}\ket{\phi(\lambda_{j})},\ \ \ N^{EC}_{ij}=\bra{\phi(\lambda_{i})}\ket{\phi(\lambda_{j})}.
	\end{aligned}
\end{equation}

In addition, by introducing complex scaling $r\rightarrow re^{i\theta}$ (or equivalent $k\rightarrow ke^{-i\theta}$), EC can also be extended to perform the task of estimating resonant states from bound states. It is noticeable that 
the inner product should be replaced by the $c$-product,
\begin{equation}
	\begin{aligned}\label{cproduct}
		&H^{EC,\theta}_{ij}=(\phi(\lambda_{i})|H^{\theta}(\lambda_{\odot})|\phi(\lambda_{j})),\ \ \ N^{EC}_{ij}=(\phi(\lambda_{i})|\phi(\lambda_{j})).
	\end{aligned}
\end{equation}

Alternatively, we can use complex scaled Hamiltonian and real training eigenvectors as in Eq.(\ref{cproduct}) or complex scaled training eigenvectors and real Hamiltonian to extract resonant energy. 
\begin{figure}[htbp] 
	\centering
	{\includegraphics[width=0.4\textwidth]{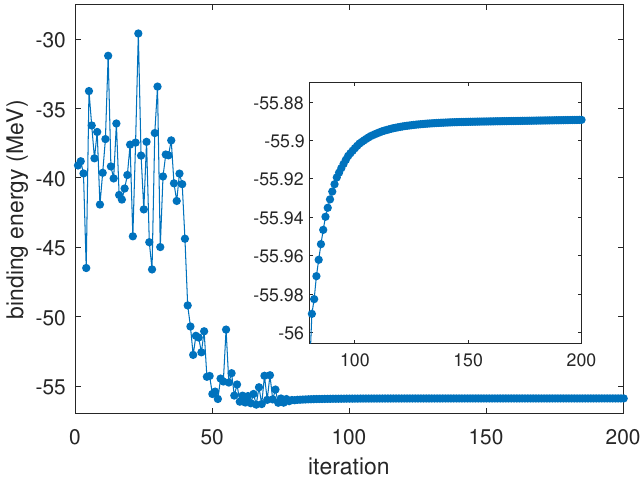} }
	\caption{The binding energy obtained through 200 iterations of the quantum neural network, with the following computational parameters: $\lambda_{\Lambda N}=1.2$, $M=0.5$. The convergent energy  is -55.89 MeV and exact energy obtained by enough basis functions is -59.47 MeV. }	\label{iteration_QNN}
\end{figure}



\section{numerical results}
\label{Numerical Results}

As a numerical test, we consider calculating the resonant $4^{+}$ state  of the ${}^{9\ }_{\Lambda}$Be system with single channel.
{\color{black}The basis vectors are produced by generator coordinates $d = 1,2,\cdots,8$ fm and $D = 1,2,\cdots,8 $ fm, requiring 6 qubits ($2^6=8\times 8 $) in QNN  to represent these 64 basis states.} 

{\color{black}
	In our eigenvector continuation implementation, variables $\lambda_{\Lambda N}$ and $M$ serve as dual parameters to compute the training data,
	\begin{equation}
		\begin{aligned}
			V_N(\lambda_{\Lambda N},M)=\lambda_{\Lambda N}V_{\Lambda N}+V_{NN}(M).
		\end{aligned}
	\end{equation}
}

The target parameters $\lambda_{\lambda N}=1$ and $M=0.573$ support a resonant $4^{+}$ state  with complex energy being $4.08-0.051i$ MeV, which is determined by using complex scaling method with different complex angles $\gamma$. Subsequent calculations are all centered around this resonant state. 
First, for eigenvector continuation with complex scaling (EC with CSM), we choose a set of parameters that ensure the system is bound, allowing us to obtain several ground-state wave functions 
$\phi_T=\{\phi_i, i=1,2\cdots\, i_{max}\}$.

\begin{figure}[htbp] 
	\centering
	{\includegraphics[width=0.525\textwidth]{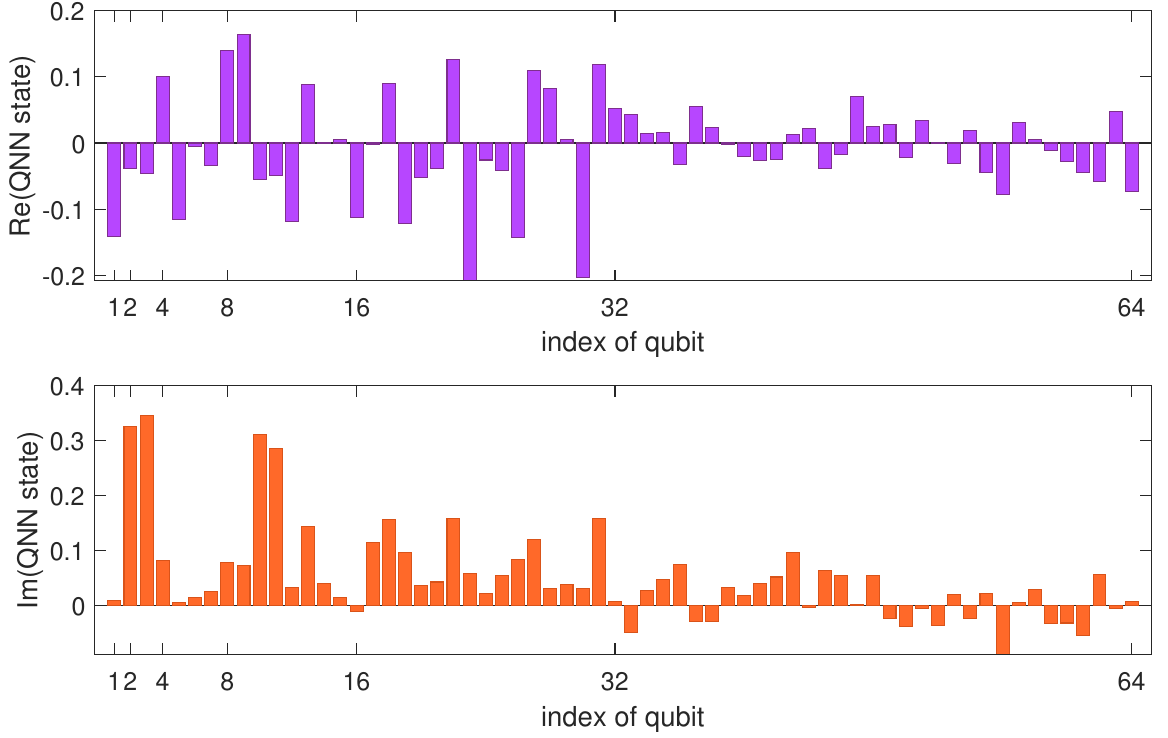} }
	\caption{The wave function (basis superposition coefficients) computed via the QNN is shown above, with the real part of the results displayed in the upper panel and the imaginary part in the lower panel. This wave function corresponds to a bound state with the parameters: $\lambda_{\Lambda N}=1.2$, $M=0.5$ and  the corresponding energy  is -55.89 MeV. }	\label{qstate}
\end{figure}
{\color{black}Notably, due to the characteristic of EC, we do not strictly require the QNN to  converge exactly to the lowest energy when iteratively evaluating the bound-state wave function. In other words, the precision demand for the wave function is relaxed. For instance, if the energy stabilizes after a certain number of iterations but further optimization demands excessive computational resources—or if the QNN’s circuit depth is insufficient to reach the true ground state—we may halt the iteration and directly employ the "inaccurate wave function." Below, we demonstrate this using the single-channel $4^{+}$ state of ${}^9_{\Lambda}$Be with the following four set of training parameters as examples,}
\begin{equation}
	\begin{aligned}
		&\{\lambda_{\Lambda N}, M\}=\{1.2 0.50\},  \{1.2 0.45\}, \{1.4 0.50\},  \{1.4 0.45\} .	\\
	\end{aligned}
\end{equation}	
As previously discussed, the generalized eigenvalue problem similarly yields a $C$-matrix compatible with the IHHL algorithm,
\begin{equation}
	\begin{aligned}\label{complexCmatrix}
		&C_{ij}(E,\beta)=(\phi|\dfrac{N^{-1}H(\gamma)-(E-\beta)}{\beta}|\phi),\ \ \  \beta\neq0.
	\end{aligned}
\end{equation}

In general when using the complex scaling method it is necessary to compute the eigenenergies corresponding to different complex scaling angles $\gamma$ to extract the final resonant energy $E_{res}$ by so-called stabilization condition. Therefore, in numerical results we take enough complex scaling angle $\gamma_i,\ (i=1,2,\cdots) $ to guarantee that the optimal angle $\gamma_{opt}$ is included.

\begin{figure}[htbp] 
	\centering
	{\includegraphics[width=0.4\textwidth]{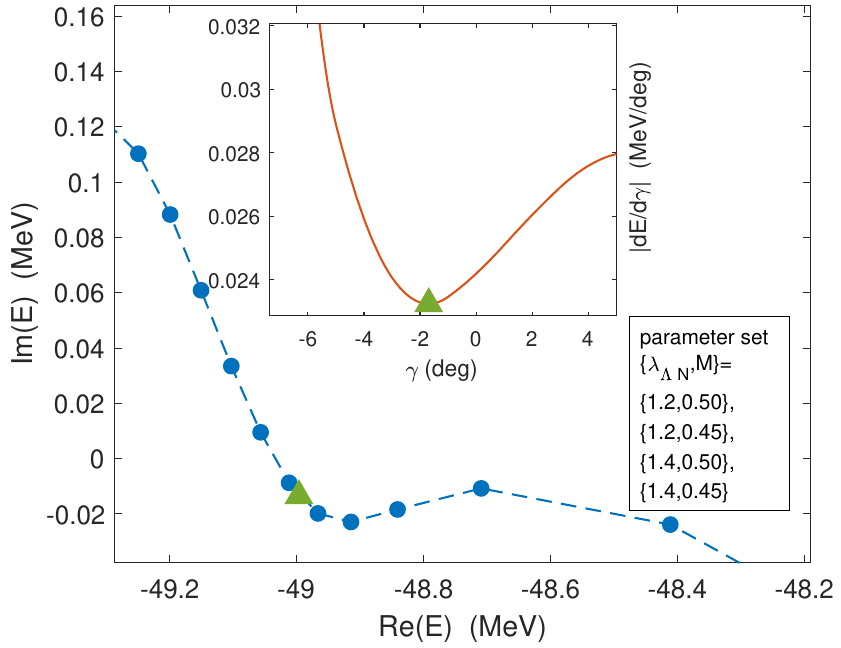} }
	\caption{The calculated resonant energies of the $4^+$ state (single channel) at different complex-scaling angles with four basis functions. The inset displays the rate of change of resonant energy versus angle, with triangle marker indicating the position of minimal rate change. Corresponding resonant energy at this angle is similarly marked. The parameters generating the eight basis vectors are also listed in the figure.}	\label{complex_spectrum_4_basis}
\end{figure}

{\color{black}As an example, for the parameter set $\{\lambda_{\Lambda N}, M\}=\{1.20, 0.50\}$, Figure \ref{gate_parameter_QNN} illustrates the evolution of quantum gate parameters during QNN iterations. The 3×6 panel represents a 3-layer circuit depth with 6 qubits, where $\theta_1$, $\theta_2$, $\theta_3$  in each subplot correspond to the rotation angles of the $RZ(\theta_1)$, $RY(\theta_2)$, $RZ(\theta_3)$ gates respectively for a given qubit and layer. Due to the periodicity of rotation angles, all quantum gate parameters are constrained to the interval $[0, 2\pi)$.  
The results demonstrate that all quantum gate parameters converge to stable values after 200 iterations, yielding a ground state energy of -55.89 MeV ( the evolution of the ground-state energy as a function of iteration number is shown in Fig.\ref{iteration_QNN}), which is slightly higher than the exact value of -59.47 MeV (the relative error is approximately $6\%$)}

{\color{black}Additionally, Figure \ref{qstate} displays the quantum state generated by the QNN with parameter set $\{ \lambda_{\Lambda N}, M\}=\{1.20,0.50\}$.} {\color{black}The upper and lower panels in Fig.\ref{qstate} share a common horizontal axis representing all $2^6=64$ possible computational basis states (from $\ket{000000}$ to $\ket{111111}$⟩) indexed by their binary values in ascending order (from 0 to 63). The vertical axes display the real and imaginary parts of the bound-state wave function's projection amplitudes onto these basis states, respectively.}

{\color{black}For other EC parameter sets, the corresponding quantum states (wave functions) can be computed using the same methodology. These wave functions will subsequently be employed to construct smaller EC matrices, which serve as inputs for the IHHL algorithm.}

{\color{black}Increasing the basis size in eigenvector continuation (EC) can improve the accuracy of resonant energy calculations. However, our focus here is not on this aspect, but rather on addressing generalized eigenvalue problems using the HHL algorithm.}
{\color{black} Therefore we will only utilize 4 basis functions above and the complex scaling angle $\gamma=-2^{\circ}$ (the integer angle closest to the interpolated optical angle shown in Fig.\ref{complex_spectrum_4_basis}) as an validation example. Here, EC based new overlap matrix $N$ and Hamiltonian matrix $H$ have been constructed using 4 basis vectors (details can be found in Appendix) and transferred into the IHHL algorithm as main ingredients.}

\begin{figure}[htbp] 
	\centering
	{\includegraphics[width=0.45\textwidth]{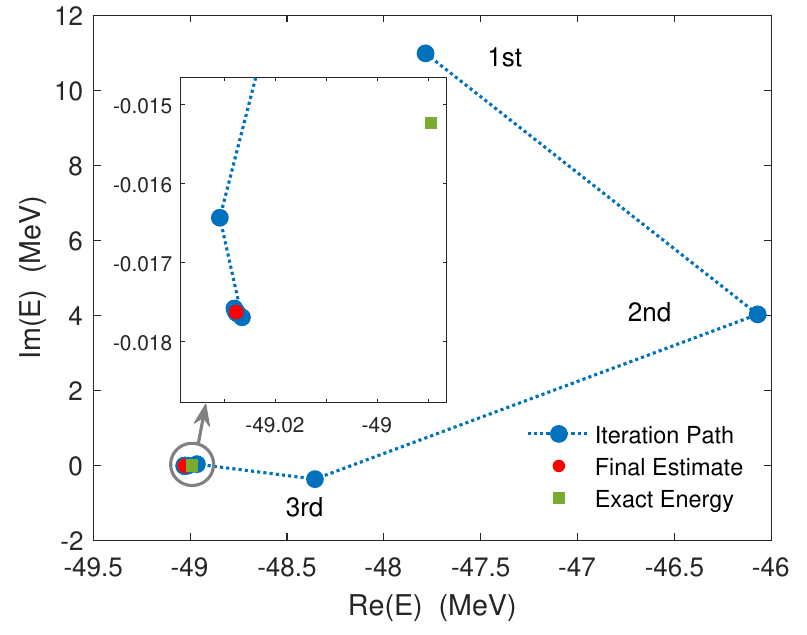} }
	\caption{The calculated resonant energies of the $4^+$ state (single channel) at  complex scaling angle $\gamma=-2^{\circ}$ with four basis functions. Parameter $\beta$ in Eq.(\ref{complexCmatrix}) is taken as 1.}	\label{HHL_iteration}
\end{figure}

\begin{figure}[htbp] 
	\centering
	{\includegraphics[width=0.5\textwidth]{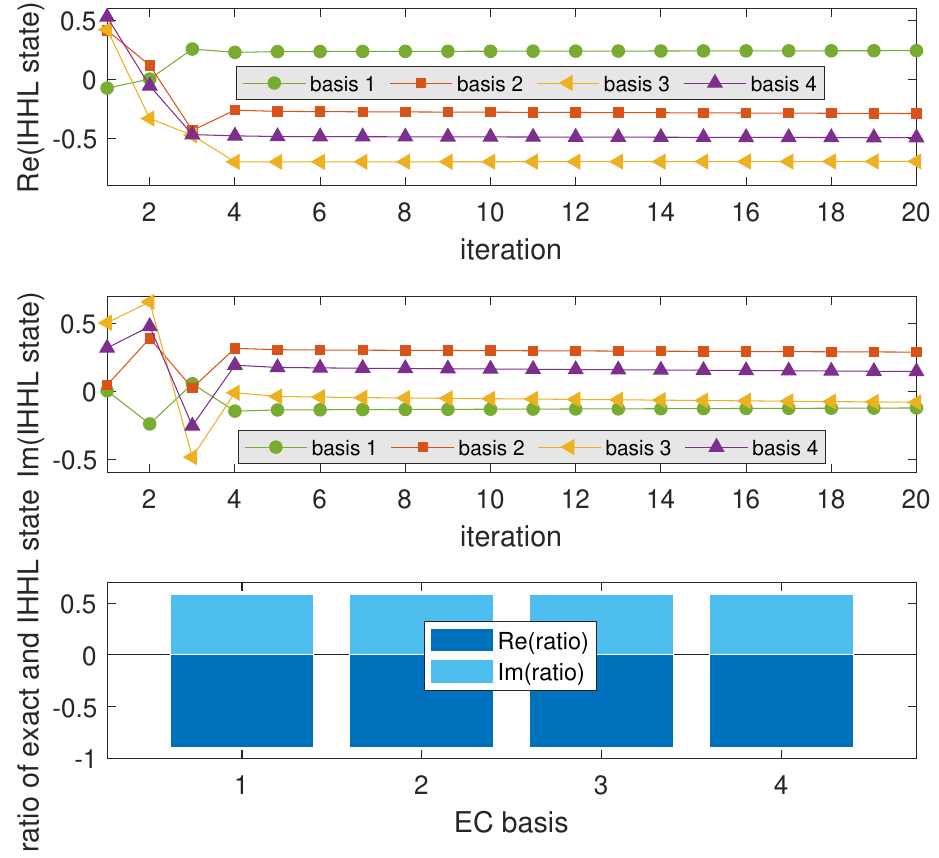} }
	\caption{The calculated resonant state with IHHL algorithm. The top and middle panels display the real and imaginary parts of the resonant state wave function (superposition amplitudes of basis vectors), respectively. The bottom panel shows the ratio between the exact wave function obtained from direct diagonalization and the IHHL results. It is clear that the closer these ratios are across different EC basis states, the more reliable and accurate the IHHL-derived resonant wave functions become.}	\label{HHL_state_iteration}
\end{figure}

In the quantum simulation the QNN-related modules can be conveniently  implemented across various quantum programming packages and  HHL algorithmic interface provided by the pyqpanda of Origin's quantum \cite{OriginalQuantum} are used in this work, which allows for the input of the Hermitian matrix $A$ and the real vector $b$.


The first eigenenergy obtained using the  IHHL algorithm are shown in Fig. \ref{HHL_iteration},  
where blue circles represent the results obtained with IHHL and the eigenvalues obtained by direct diagonalization of the matrix $H ^{\theta=-2^{\circ}}$ are marked with green square. After few iterations the first eigenvalues converge to our given accuracy and the final estimate is marked by red circle. 
It can be obviously shown that with IHHL algorithm the converged eigenenergies will be obtained after a few iterations, which demonstrates its high computational efficiency.

It should be noted that, the initial wave function we choose here (listed in the appendix) is such that the target resonant state is already obtained in the first solution, otherwise we will continue to implement the  projection method to ensure orthogonality and thus avoid convergence of the results to the  already solved eigenstates. Of course, the order of the eigenvalue solution may change if different initial wave functions are used. 

In addition, Figure \ref{HHL_state_iteration} presents the convergence behavior of the first eigen wave function. The upper and middle panels demonstrate the convergence of the eigenvector's real and imaginary components during IHHL iterations, showing stabilization within approximately ten iterations - consistent with the eigenenergy convergence. The lower panel displays the ratio between the exact eigenvector and our 20th-iteration result, where dark and light blue regions represent the ratio's real and imaginary parts respectively. The near-constant values across all four components quantitatively validate the reliability of our obtained eigenvectors. At this stage we have investigated a resonant state in hypernucleus ${}^9_{\Lambda}$Be with QNN-IHHL framework, this fully quantum algorithm can be directly extended to study resonance phenomena across diverse nuclear systems - from single-$\Lambda$ and double-$\Lambda$ hypernuclei to other complex multi-cluster nuclear structures. This consistent methodology may also show promising potential for ab initio calculations of exotic nuclear states and  offering enhanced computational capabilities.

\section{conclusion}
\label{Conclusions}
In this work, we propose a novel quantum algorithm QNN-IHHL for solving few-cluster resonant states and develop a comprehensive framework of full-process quantum algorithms. Based on the original application of IHHL algorithm in studying non-Hermitian systems  we innovatively introduce the QNN framework,  whose parameterized circuits provide trainable function approximations and enable efficient representation of high-dimensional quantum states. Additionally, by combining complex scaled eigenvector continuation with iteration strategy based on HHL algorithm, our approach efficiently  extends the quantum computational capabilities from bound states to resonant states.   Taking the $4^{+}$ resonance of the ${}^9_{\Lambda}$Be system as an example, with the help of quantum simulations we obtain the resonant energy consistent with the conventional method. Our future work will focus on experimental realization using near-term quantum devices and applications to more complex nuclear systems.

\section*{Acknowledgements}
This work is supported by the National Natural Science Foundation of China (Grants No.\ 12035011,  No.\ 124B2100, No.\ 11905103, No.\ 11947211, No.\ 11961141003, No.\ 12022517, No.\ 12375122 and No.\ 12147101), by the National Key R\&D Program of China (Contracts No.\ 2023YFA1606503)


\begin{appendix}
\begin{widetext}

\section{}
In the following, we list the quantum circuit parameters obtained using the QNN for the four EC basis wave functions (corresponding to parameter sets $\{\lambda_{\Lambda N}, M\}=\{1.2 0.50\},  \{1.2 0.45\}, \{1.4 0.50\},  \{1.4 0.45\}$ respectively),

\begin{equation}
	\begin{bmatrix}
		6.2777443 & 2.9623668 & 4.1739912 \\
		2.1441712 & 3.2506409 & 4.0256934 \\
		3.3965152(-5) & 6.0410719 & 4.7356915 \\
		1.8482275 & 5.6914926 & 4.8712077 \\
		1.1949457(-13) & 3.5505292 & 1.0660791 \\
		1.9448582 & 5.4494443 & 1.8766041
	\end{bmatrix}
	,
	\begin{bmatrix}
		3.2506917 & 0.33835053 & 5.0115643 \\
		2.2775216 & 0.28397676 & 5.6699934 \\
		4.8914018 & 3.1740475 & 3.5373149 \\
		0.57324088 & 5.7461190 & 5.5679650 \\
		2.6023750 & 5.8771977 & 0.71344692 \\
		5.0057235 & 1.2177472 & 3.3084846
	\end{bmatrix}
	,
	\begin{bmatrix}
		3.6596761 & 5.3445787 & 3.3469312 \\
		5.5604734 & 1.3144962 & 3.1209221 \\
		0.88926297 & 4.7070713 & 5.2865720 \\
		3.6204617 & 5.8235817 & 1.9200119 \\
		0.60789013 & 1.1107392 & 6.2551217 \\
		2.0389147 & 4.6483450 & 0.12981953
	\end{bmatrix}
\end{equation}

\begin{equation}
	\begin{bmatrix}
		6.2777443 & 1.6532806 & 5.5299764 \\
		2.1441712 & 1.7400076 & 6.0886784 \\
		3.3965152(-5) & 6.2642851 & 2.0849619 \\
		1.8482275 & 3.1492755 & 2.8281085 \\
		5.8295873(-16) & 3.0775306 & 1.5378615 \\
		1.9448582 & 3.0582225 & 2.9575124
	\end{bmatrix}
	,
	\begin{bmatrix}
		3.0978861 & 0.88025045 & 1.2276460 \\
		2.4936917 & 0.010903232 & 0.011435008 \\
		3.7364287 & 2.5512588 & 6.0294924 \\
		4.9692655 & 3.1640944 & 0.19206865 \\
		4.0859952 & 3.6108334 & 1.2612017 \\
		5.0374656 & 4.9234509 & 0.57810557
	\end{bmatrix}
	,
	\begin{bmatrix}
		3.0744302 & 0.65630126 & 5.9030437 \\
		6.0204277 & 3.6330497 & 4.3871384 \\
		5.2626009 & 5.8649426 & 1.9436437 \\
		6.2162886 & 0.060260054 & 0.44858441 \\
		2.0183523 & 0.39174813 & 6.1193180 \\
		6.0965328 & 4.1774902 & 1.5052414
	\end{bmatrix}
\end{equation}

\begin{equation}
\begin{bmatrix}
		6.2777443 & 3.7113049 & 5.8903913 \\
		2.1441712 & 3.6780977 & 5.2731752 \\
		3.3965152(-5) & 6.1982117 & 1.7222717 \\
		1.8482275 & 2.8075969 & 2.6714938 \\
		1.4487108(-14) & 2.6655653 & 6.8465449(-2) \\
		1.9448582 & 5.2419791 & 3.8776989
\end{bmatrix}
,
\begin{bmatrix}
	0.86774749 & 3.7376420 & 0.74231458 \\
	0.66126758 & 3.5271878 & 1.6559840 \\
	1.5172551 & 3.1828408 & 4.1716714 \\
	2.0018253 & 3.1072955 & 3.5669951 \\
	0.12009725 & 3.0624554 & 0.72391278 \\
	5.8197627 & 1.9134910 & 4.6149969
\end{bmatrix}
,
\begin{bmatrix}
	0.49403602 & 0.59579062 & 6.1259356 \\
	2.8070350 & 4.6902432 & 0.46483782 \\
	3.0910778 & 5.5677624 & 5.5603318 \\
	3.7144027 & 3.1327314 & 2.1026759 \\
	2.2346506 & 1.1773512 & 5.8904195 \\
	1.4376781 & 4.9916224 & 2.8062603
\end{bmatrix}
\end{equation}

\begin{equation}
	\begin{bmatrix}
		6.2777443 & 4.6395106 & 2.0782170 \\
		2.1441712 & 0.76407343 & 2.3665769 \\
		3.3965152(-5) & 5.9719467 & 2.8865707 \\
		1.8482275 & 2.9538057 & 2.9023886 \\
		1.5688326(-14) & 0.19871604 & 1.5182071 \\
		1.9448582 & 0.098396666 & 3.5893679
	\end{bmatrix}
,
	\begin{bmatrix}
		3.7577772 & 3.6340573 & 1.5114065 \\
		3.4197664 & 5.0242534 & 1.8561713 \\
		2.6410367 & 0.39176977 & 2.4164686 \\
		5.0180240 & 1.6082621 & 3.2162220 \\
		2.0420129 & 3.3161902 & 6.0162206 \\
		0.64636427 & 4.6717138 & 3.2288892
	\end{bmatrix}
,
	\begin{bmatrix}
		1.4781995 & 5.8066125 & 3.0777385 \\
		0.51672262 & 4.1419640 & 4.9904366 \\
		3.0799878 & 6.0354042 & 0.20994452 \\
		6.1868892 & 1.5643044 & 4.6456366 \\
		0.76679182 & 0.067824759 & 0.082950279 \\
		5.7389526 & 5.1092334 & 5.7486968
	\end{bmatrix}
\end{equation}
where the three matrices in each equation represent three layers in QNN and each matrix corresponds to the parameters of 3 quantum gates $R_Z,R_Y,R_Z$ in  6 qubits. Numbers in parentheses denote scientific notation.

With these four EC basis wave  functions the Hamiltonian matrix ${H}_{\text{res}}$ and overlap matrix $N_{\text{res}}$ with complex scaling angle $\gamma=-2^{\circ}$ are computed as,
\begin{equation}
	\begin{aligned}
&H_{\text{res}}=\\
&
 \begin{bmatrix}
 	 89.55105672 + 29.96865528i & -71.61897281 - 54.24046502i &  71.32224612 + 54.5001247i  &  84.10305776 - 26.89469849i \\
 	-71.61897281 - 54.24046502i &  49.07619985 + 75.61013658i & -47.19178977 - 73.34045931i & -88.66338303 - 3.61063443i \\
 	71.32224612 + 54.5001247i  & -47.19178977 - 73.34045931i &  46.82508651 + 72.12727791i &  85.84825993 + 3.55833856i \\
 	84.10305776 - 26.89469849i & -88.66338303 - 3.61063443i &  85.84825993 + 3.55833856i &  53.3190466 - 69.41101231i
 	\notag
	\end{bmatrix}
,
	\end{aligned}
\end{equation}

\begin{equation}
	\begin{aligned}
&N_{\text{res}}=\\
&
\begin{bmatrix}
 -1.81352045 - 0.6028637i  &  1.45656633 + 1.0973103j  & -1.44743132 - 1.10790096i & -1.70943956 + 0.54708644i \\
1.45656633 + 1.0973103j  & -1.01540018 - 1.54838119i &  0.95991351 + 1.49245168i &  1.81973426 + 0.07490585i \\
-1.44743132 - 1.10790096i &  0.95991351 + 1.49245168i & -0.95030104 - 1.46887722i & -1.74893355 - 0.07627396i \\
-1.70943956 + 0.54708644i &  1.81973426 + 0.07490585i & -1.74893355 - 0.07627396i & -1.09854298 + 1.42115795i
\end{bmatrix}
.
	\end{aligned}
\end{equation}

{\color{black}
Using the matrices above and the initial wave function $\phi= [1,2,3,4]$,  the first eigenstate obtained from IHHL algorithm happens to correspond to the target resonant state. 
}

\end{widetext}
\end{appendix}
\bibliography{example}

\end{document}